\documentclass[useAMS,usenatbib]{mn2e} 
\pdfoutput=1

\usepackage[fleqn]{amsmath}
\usepackage{graphicx}
\usepackage{upgreek}
\newcommand{\half}{\frac{1}{2}}

\newcommand{\rd}{\rmn{d}}

\newcommand{\x}{{\bmath{x}}}
\renewcommand{\i}{{\rmn{i}}}

\newcommand{\vdot}{{\bmath{\cdot}}}
\newcommand{\grad}{\bmath{\nabla}}
\newcommand{\B}{{\bmath{B}}}

\newcommand{\thth}{\hspace{1.5pt}}

\newcommand\Div{\grad\vdot\thth}

\newcommand{\bv}{Brunt-V\"ais\"al\"a}

\newcommand{\mnras}{MNRAS}
\newcommand{\apj}{ApJ}

\newcommand{\apjl}{ApJL}

\newcommand{\solphys}{Solar Phys.}
\newcommand{\nat}{Nature}

\begin{document}

%\end{document}

\title[Phase Shifts in Local Helioseismology]{Magnetic and Thermal Phase Shifts in the Local Helioseismology of Sunspots}

\author[P.~S.~Cally]{Paul S. Cally\thanks{Email: paul.cally@sci.monash.edu.au}\\
Centre for Stellar and Planetary Astrophysics,
School of Mathematical Sciences,
Monash University, Victoria, Australia 3800}

\maketitle

\begin{abstract}
Phase perturbations due to inclined surface magnetic field of active region strength are calculated numerically in quiet Sun and simple sunspot models in order to estimate and compare the direct and indirect (thermal) effects of the fields on helioseismic waves. It is found that the largest direct effects occur in highly inclined field characteristic of penumbrae, and scale roughly linearly with magnetic field strength. The combined effects of sunspot magnetic and thermal anomalies typically yield negative travel-time perturbations in penumbrae. Travel-time shifts in umbrae depend on details of how the thermal and density structure differs from the quiet Sun. The combined shifts are generally not well approximated by the sum of the thermal and magnetic effects applied separately, except at low field strengths of around 1 kG or less, or if the thermal shift is small. A useful rule-of-thumb appears to be that travel-time perturbations in umbrae are predominantly thermal, whereas in penumbrae they are mostly magnetic.
\end{abstract}


%%%%%%%%%%%%%%%%%%%%%%%%%%%%%%%%%%%%%%%%%%%%%%%%%%%%%
\section{INTRODUCTION}
Time-Distance helioseismology (TD; \citealt{duvall93}) and Helioseismic Holography (HH; \citealt{lb00}) are prominent examples of local helioseismic techniques which rely on accurate identification of oscillation phase in order to probe the solar interior. Active regions, especially sunspots, have been particular targets of these methods in an ongoing quest to better understand the Sun's magnetic activity \citep{kos00,bl00}. However, \cite{lb05} and \cite{sbcl05} draw particular attention to the ``showerglass'' effect, produced by surface phase perturbations, which obscures the subsurface holographic image, particularly in penumbrae. \cite{cr07} also identify ``surface magnetic contamination'' of TD data in penumbrae, and \cite{bb08} emphasize the frequency dependence of travel times.

Recently, \cite{cally08} employed exact solutions for waves in a two-dimensional (2D) isothermal model with uniform inclined magnetic field to firmly identify near-surface phase shifts in helioseismic waves. These occur in particular at the upper turning point of fast (magnetically dominated) waves, and at the mode conversion level $z_\mathrm{eq}$ (where the Alfv\'en speed $a$ and the sound speed $c$ coincide). This is a purely magnetic effect; no thermal perturbations were introduced. Such direct magnetic effects are clearly prime candidates for explaining the observed surface anomalies.

Unfortunately, the exact solution approach is restricted to the isothermal case, and to 2D, and so is more a proof of concept rather than a directly applicable model in the realistic solar context. In this paper we relax these restrictions, and quantitatively calculate the surface phase shifts due to both 3D magnetic and 1D thermal effects, individually and in combination.

%%%%%%%%%%%%%
\subsection{Effects of Magnetic and Thermal Anomalies}
It is common practice in local helioseismology to linearize the effects on the travel time perturbations. For example, \cite{kd97} propose the following expression for the travel-time perturbation in the ray approximation
\begin{multline}
\Delta\tau(\x_1,\x_2) = \\
- \int_\Gamma
\frac{\hat{\mathbf{n}}\vdot\mathbf{U}}{c^2}
+ \frac{1}{v_p}\,\frac{\delta c}{c}
+\frac{\delta\omega_c}{\omega_c}\,\frac{\omega_c^2v_p}{\omega^2 c^2}
+\half \left(\frac{a^2}{c^2}-\frac{(\mathbf{a\vdot k})^2}{c^2k^2}\right)\,\rd s\, ,            \label{dtau}
\end{multline}
where $\Gamma$ is the ray path joining the two surface points $\x_1$ and $\x_2$, $\hat{\mathbf{n}}$ is the unit vector along $\Gamma$, $c$ and $a$ are the sound and Alfv\'en speeds, $\mathbf{a}$ is the Alfv\'en velocity, $\mathbf{U}$ is the underlying flow speed, $\omega$ is the frequency, $\mathbf{k}$ is the wavevector with $k=|\mathbf{k}|$ the wavenumber, $\omega_c$ is the acoustic cut-off frequency, and $v_p=\omega/k$ is the phase speed. The magnetic field is being treated as a perturbation to the background non-magnetic state (hence the lack of a $\delta \mathbf{a}$ term), with $\delta c$ and $\delta\omega_c$ characterizing the thermal perturbation. Fermat's Principle -- that the travel time varies only at second order with perturbations to the path $\Gamma$ -- is invoked to avoid a $\delta\Gamma$ contribution. Although lacking an accounting of the effect of the {\bv} (buoyancy) frequency, and incorporating a description of the magnetic influence which is valid only if $a\ll c$, equation (\ref{dtau}) illustrates how $\Delta\tau$ is perceived as depending additively on flow (first term), thermal (second and third terms), and magnetic (final term) effects. 

However, \cite{w62} proves that Fermat's Principle applies within ray theory only if the dispersion function $D(\omega,\mathbf{k})$ is homogeneous in $\omega$ and $\mathbf{k}$, \emph{i.e.}, $D(\alpha\omega,\alpha\mathbf{k})=\alpha^m\,D(\omega,\mathbf{k})$ for some $m$, which is not the case if acoustic cut-off or {\bv} frequencies are included \citep[see also][]{bc01}. Furthermore, classical ray theory breaks down in the neighbourhood of the mode conversion layer $z\approx z_\mathrm{eq}$, where rays are split and their subsequent paths radically altered \citep{sc06,cally07}, again invalidating Fermat's Principle. Surface magnetism in active regions certainly cannot be regarded as a mere linear perturbation.

In practice, only flow and thermal perturbations have been incorporated into TD inversions up till now. Recently though, it has become clear that direct magnetic effects, through both magnetic perturbations to the wave speeds and to the ray paths, are crucial, indeed dominant, in the surface layers of sunspots
\citep{cgd08,mc08,mhc08}.\footnote{Both \cite{mc08} and Robert Cameron (private communication) find that travel-time perturbation magnitudes are greatly reduced in most cases if direct magnetic effects are artificially ``turned off'' whilst retaining the thermal perturbations in sunspot models, strongly suggesting that the direct effects generally dominate.} In particular, based on sophisticated numerical simulations, \citeauthor{cgd08} conclude that ``constraining the sunspot model with helioseismology is only possible because the direct effect of the magnetic field on the waves has been fully taken into account''. 

A further issue with current practice arises from equation (\ref{dtau}) or similar: whereas inversion is based on such equations, the ``observed'' $\Delta\tau$, which comprises the data being inverted, derives from measured phase shifts. However, as \cite{cally08} demonstrates, phase is not continuous along a ray path, but instead suffers significant jumps at turning and mode conversion points. It may be hoped that, although total travel time measured through phase will certainly not agree with that found by integrating phase speed along ray paths, the perturbations may be tolerably accurate. But this is by no means certain when strong surface magnetic fields are involved.

Mean one-way travel-time perturbations in the range $-50\mathrm{\,s}\la\Delta\tau\la40\mathrm{\,s}$ are reported by \citet{cbk06}, with negative $\Delta\tau$, for deep, comparatively low-$\ell$ modes, and positive $\Delta\tau$ for shallow high-$\ell$ waves. Using ridge-filtering rather than phase-speed filtering, \citet{cr07} and \citet{bb08} find that positive travel-time perturbations are more prevalent for shallow waves in sunspot umbrae, with negative shifts more characteristic of penumbrae. However, they show that such details are very sensitive to the phase-speed and frequency filters used, with more extensive negative-$\Delta\tau$ regions at higher frequencies (4.5 mHz).

With these points understood, the specific questions to be addressed here are:
\begin{enumerate}
\item What ``travel-time'' shifts does the magnetic field alone produce (always bearing in mind that these are conventionally calculated from the observed phase shifts using $\Delta\tau=-\omega^{-1}\Delta\varphi$, and therefore are not strictly travel-time shifts at all when there are phase jumps)?
\item What travel-time shifts does the thermal structure of a sunspot produce, in the absence of magnetic field?
\item What is $\Delta\tau$ due to the combined thermal and magnetic anomalies, and is it a simple sum of the individual effects?
\end{enumerate}

%%%%%%%%%%%%%%%%%%%%%%%%%%%%%%%%%%%%%%%%%%%%%%%%%%%%%

\begin{figure}
\begin{center}
\includegraphics[width=.85\hsize]{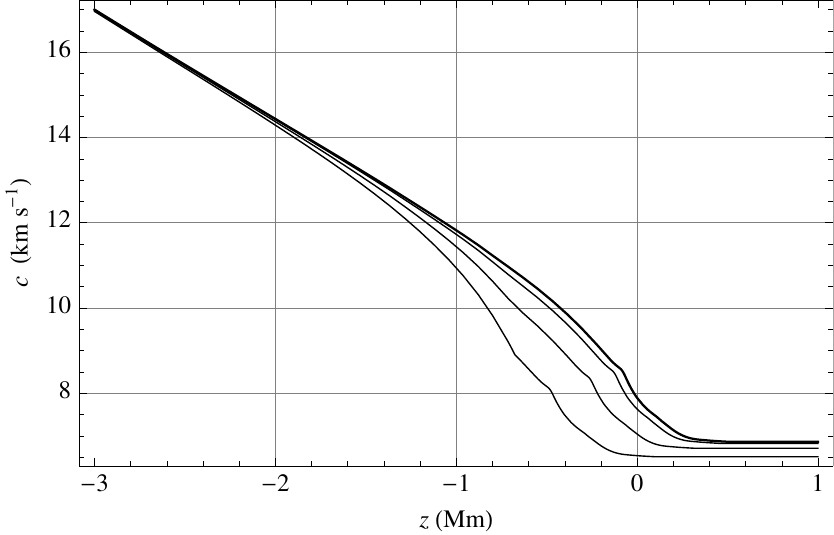}
\caption{The sound speed profiles near the surface in the MMMS models with (top to bottom) 0 kG, 1 kG, 2 kG, and 3 kG magnetic fields respectively.}
\label{fig:MMMS}
\end{center}
\end{figure}

\section{MODEL}
The model is essentially that of \cite{cg08}. The well-known quiet Sun Model S of \cite{S} is supplemented by a uniform magnetic field $\B=B\,(\sin\theta\cos\phi,\sin\theta\sin\phi,\cos\theta)$ inclined at angle $\theta$ to the vertical and angle $\phi$ to the $x$-$z$ Cartesian plane in which the wave is defined to propagate. Since the field is uniform, there is no net magnetic force on the medium, so the model is formally consistent. The 2D case corresponds to $\phi=0$. An infinite isothermal atmosphere is placed above Model S ($z>0.5$ Mm).
The thermal anomaly is crudely accounted for in the Magnetically Modified Model S (MMMS, see figure \ref{fig:MMMS}) of \citet[Appendix A]{sc06}, which reduces the Model S gas pressure by the constant magnetic pressure, and hence leaves the density unchanged (except in the overlying atmosphere where this pressure balance is untenable). The MMMS models take field strength $B$ as a parameter, and reduce the temperature near the surface compared to Model S, with greater cooling as $B$ increases. Although ad hoc, the MMMS models mimic the crucial cool surface layer of a real sunspot, and will be adequate to test the relative roles of the thermal and magnetic anomalies. An alternate thermal model which allows for the atmosphere to compress downward with decreasing temperature is briefly discussed in Section \ref{sec:comb}.

It has become abundantly clear in recent years that sunspots are a very shallow phenomenon, with most of their helioseismic impact being restricted to within a few hundred kilometres of the surface. The density and Alfv\'en scale heights are similarly short. This is to be compared with horizontal distances of several megametres or more over which the field and thermal structure vary. It is therefore appropriate to explore horizontally homogeneous models in the first instance, especially when concerned with waves which approach the surface nearly vertically, and resolve the umbra and penumbra.

The approach is that of a scattering experiment. We envisage an infinite horizontal driving plane at depth $z=z_\mathrm{b}<0$ injecting fast magneto\-acoustic waves with $\exp[\i(k_x\,x-\omega\, t)]$ horizontal and temporal dependence into the overlying near-photosphere. The horizontal wavenumber $k_x$ is chosen so that the wave's lower turning point would be at a prescribed depth, $z_1 < z_\mathrm{b}$, if the model were not truncated at $z_\mathrm{b}$. Typically, $z_\mathrm{b}$ is placed deep enough that $a\ll c$, and the fast and slow magneto\-acoustic waves are well-decoupled. Since the sound speed dominates in this region, fast waves are overwhelmingly acoustic, with at most only minor magnetic characteristics. The upgoing fast (acoustic) wave is partially converted to a fast (magnetic) wave near $z_\mathrm{eq}$: mode conversion in the 2D case, where the Alfv\'en wave decouples, is extensively explained using generalized ray theory in \cite{sc06} and \cite{cally07}. Above $z_\mathrm{eq}$ it refracts from the steep Alfv\'en speed gradient and propagates back downward. The remainder of the incident wave's energy is transmitted as a nearly field-aligned slow (acoustic) wave, and, in the 3D case, as an Alfv\'en wave. On its way back downward, the fast (magnetic) wave again meets $a=c$ and is again partially converted to a fast (acoustic) wave, partially to an Alfv\'en wave, and partially (transmitted) to a slow (magnetic) wave. It is the downgoing fast wave in $z < z_\mathrm{eq}$ that is of prime interest here, as it is this wave that re-enters the helioseismic wave field. It skips in the usual manner, and may be observed and interrogated at the surface in the surrounding quiet Sun. 

There is another route to the skipping fast wave though. If the wave frequency $\omega$ is less than the ramp-reduced acoustic cutoff frequency $\omega_c\cos\theta$, the transmitted slow (acoustic) wave in $z>z_\mathrm{eq}$ will reflect, pass through $a=c$, and partially transmit to a fast (acoustic) wave in $z<z_\mathrm{eq}$, where it will join the fast wave which took the first route.

We are interested in both the downgoing fast wave's energy flux (relative to that of the original injected wave) $\mathcal{R}\in[0,1]$ and its phase. Using numerical solution of the full linearized $6^\mathrm{th}$-order wave equations and a least-squares fit of the calculated acoustic field $\Psi=\rho^{1/2}c^2\,\Div\bxi$ to the standard WKB acoustic solution, $\Psi=C_+\, f_+(z) + C_-\,f_-(z)$, the upgoing and downgoing coefficients $C_+$ and $C_-$ are found (see the appendix for details). Here, $\rho$ is the density and $\bxi$ is the plasma displacement. The fit is performed over a selected interval $z_\mathrm{b}<z<z_\mathrm{s}$ chosen to sit comfortably within the acoustic cavity, and its accuracy is normally excellent in practice. (It does become more difficult though, and the results slightly more uncertain, for a few of the very shallow waves with lower turning point $z_1=-3$ Mm explored in Section \ref{sec:3D}, but the uncertainties are more matters of detail than they are fundamental.)\phantom{.} Then, having defined $f_+$ and $f_-$ to be equal at $z_\mathrm{b}$, the reflection coefficient and total phase change at $z_\mathrm{b}$ are
\begin{equation}
\mathcal{R}= \left| \frac{C_-}{C_+} \right|^2 \qquad \mbox{and} \qquad
\delta\varphi = -\arg\left(\frac{C_-}{C_+}\right)\, .  \label{R-delta}
\end{equation}

The raw $\delta\varphi$ is not very useful in this form. It can be altered by simply moving $z_\mathrm{b}$ up or down. However, the significant quantity is the difference between $\delta\varphi$ in the magnetic and non-magnetic cases, $\Delta\varphi=\delta\varphi-\delta\varphi_0$, both calculated with the same $z_1$. This is independent of $z_\mathrm{b}$, provided it is well within the $c\gg a$ regime. Unlike \cite{cally08}, we do not calculate the individual phase jumps at the conversion and top turning points, but only the total perturbation through the surface layers. The total inferred phase travel time perturbation, as used in TD, is then $\Delta\tau=-\omega^{-1}\Delta\varphi$: phase advance $\Delta\varphi>0$ corresponds to negative $\Delta\tau$, \emph{i.e.}, faster propagation.

Three boundary conditions are specified at the top, $z_\mathrm{t}$, situated in the overlying isothermal atmosphere: that the fast wave is evanescent rather than exponentially growing with height; that the slow wave is either upwardly propagating or evanescent, according as $\omega$ exceeds $\omega_c\cos\theta$ or \emph{vice versa}; and that the Alfv\'en wave is outgoing. \cite{cg08} present full details of how these conditions are imposed. In the spirit of a scattering experiment designed to probe the surface layers, we do not attempt to model the chromosphere. The isothermal layer, in which we have exact solutions, is a convenient ``observation region'' in which the transmitted waves may be examined. Further reflection off the rise from temperature minimum to chromospheric temperatures, or off the transition region, as well as radiative effects, will produce somewhat altered results, but they are outside our focus of interest.

The mathematical problem is completed by the further imposition of three boundary conditions at the bottom, $z_\mathrm{b}$: fast, slow and Alfv\'en waves are outgoing, and an incoming fast (acoustic) wave is imposed. No restriction is placed on the resulting outgoing fast wave, of relative flux $\mathcal{R}$.

%%%%%%%%%%%%%%%%%%%%%%%%%%%%%%%%%%%%%%%%%%%%%%%%%%%%%
\section{RESULTS}

\begin{figure}
\begin{center}
\includegraphics[width=.85\hsize]{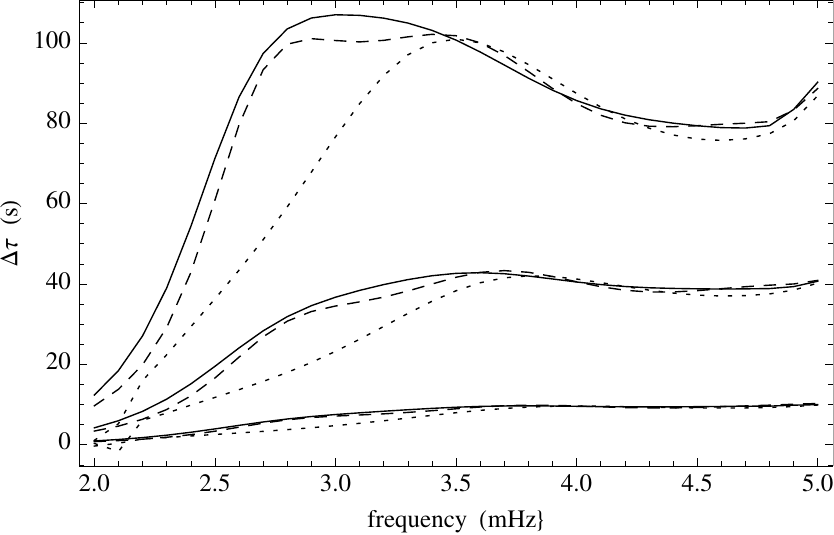}
\caption{Two-way travel-time perturbations as a function of frequency for the MMMS model with magnetic field suppressed, but its thermal consequences retained. The top three curves correspond to $B=3$ kG, the middle three to 2 kG, and the bottom three to 1 kG. The full curves are for waves with bottom turning points at $z_1=-15$ Mm, the dashed curves for $z_1=-10$ Mm, and the dotted curves for $z_1=-5$ Mm.}
\label{fig:thermal}
\end{center}
\end{figure}

\begin{figure}
\begin{center}
\includegraphics[width=\hsize]{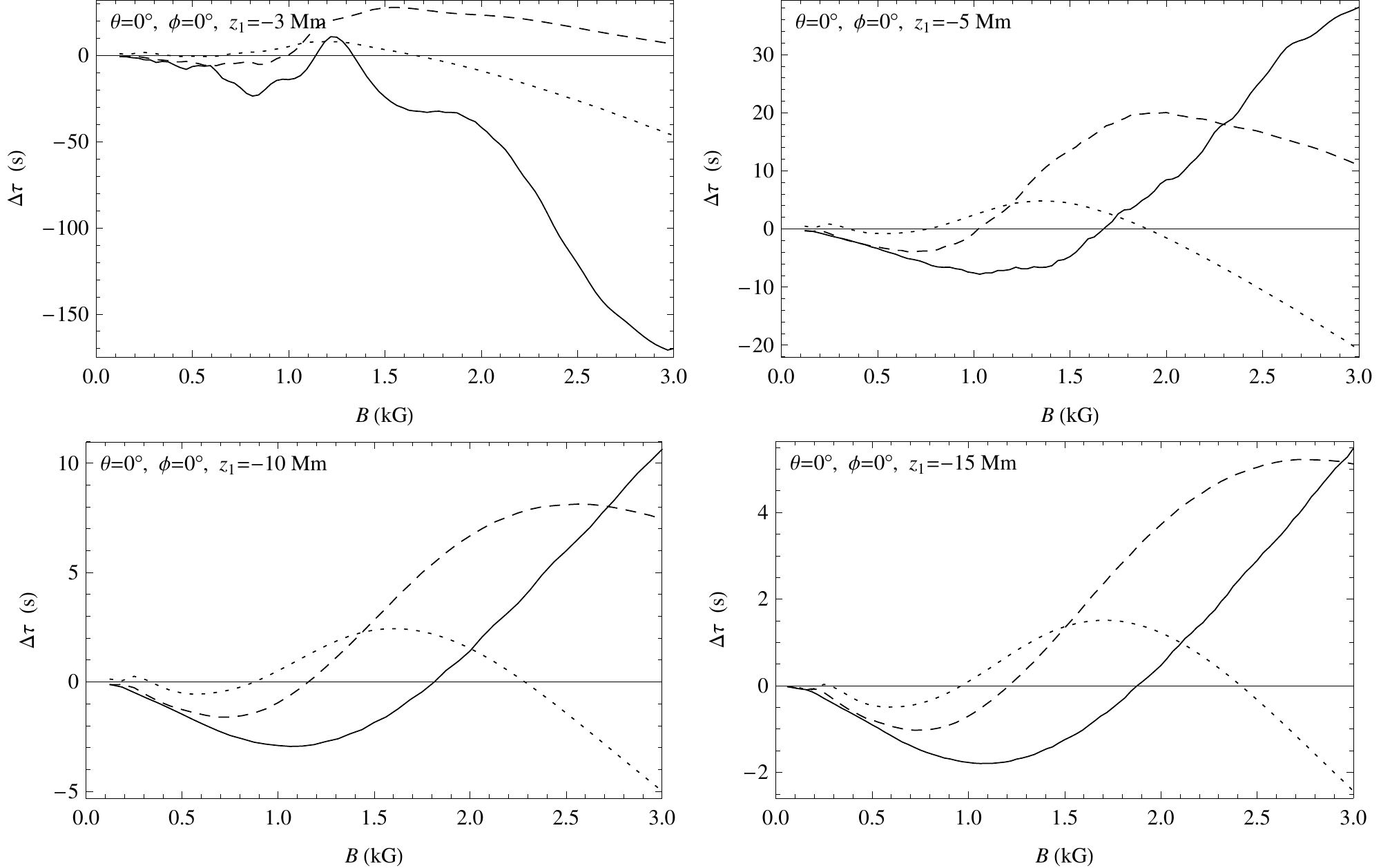}
\caption{Travel-time perturbation $\Delta\tau=-\omega^{-1}\Delta\varphi$ (seconds) for vertical magnetic field and four lower turning depths $z_1$ (labelled) as a function of magnetic field strength (kG). Full curves: 3 mHz; dashed curves: 4 mHz; dotted curves: 5 mHz.}
\label{fig:B}
\end{center}
\end{figure}

\subsection{Thermal Anomaly Only}  \label{sec:thermonly}
Figure \ref{fig:thermal} displays the travel-time perturbations for the MMMS models with thermal perturbations corresponding to $B_\mathrm{eff}=1$, 2, and 3 kG effective magnetic fields, but with the direct magnetic effects on the waves suppressed. For comparison with the observed travel-time perturbations \citep{cbk06} mentioned earlier, note that those presented here are \emph{two-way} perturbations; \emph{i.e.}, the wave has travelled both up and down from and to $z_b$, and so is double the one-way shifts. The cooler models have produced \emph{increased} travel times, with greater effect for larger $B_\mathrm{eff}$, and lesser at low frequencies, at which the wave turns over much lower, and hence does not substantially sample the surface thermal perturbation.
This is a feature of the MMMS models, in which temperature is reduced but density is unchanged from model S. Hence, the acoustic cutoff frequency $\omega_c=(c/2H)\sqrt{1-2H'}$ (where $H$ is the density scale height) is also reduced, and if anything the wave reaches higher before being reflected. The combination of lower sound speed and longer path length naturally results in positive travel-time shifts.

These results are not definitive. Different thermal models can produce quite different timings. For comparison, an alternate thermal model with reduced surface density and \emph{negative} non-magnetic travel-time perturbations is briefly addressed in Section \ref{sec:comb}. However, the point here is to test whether thermal and magnetic effects are additive, rather than to produce a fully realistic sunspot model.

%%%%%%%%%%%%%%%%%%%%%%%

\subsection{Magnetic Anomaly Only} \label{sec:magonly}
The effects of a uniform magnetic field superimposed on unmodified Model S are examined next. These are presented in detail, as a major theme of this paper is to emphasize the direct magnetic effects which are too-often ignored in local helioseismology.

%%%%%%%%%%%
\subsubsection{Vertical Magnetic Field}
Near-vertical magnetic field of around 3 kG or more may be found in large sunspot umbrae. Figure \ref{fig:B} presents the \emph{inferred} travel-time shift $\Delta\tau=-\omega^{-1}\Delta\varphi$ as a function of field strength for several frequencies and lower turning depths. (Note though that $\Delta\tau$ is simply a measure of phase shift $\Delta\varphi$, and is not necessarily a true travel-time shift.)\phantom{.} As expected, $\Delta\tau$ diminishes with reducing $B$, though the limit cannot be approached indefinitely for numerical reasons.\footnote{The governing differential equations for the magnetic case are either fourth order (2D) or sixth order (3D), whereas the non-magnetic equation is second order (see Appendix \ref{app:A}). The three cases therefore require different numerical solution processes, with different numbers of boundary conditions. As $B\to0$, the slow and Alfv\'en wavelengths vanish, so letting $B$ become very small requires ever decreasing step-lengths, which of course cannot be continued through to the limit.} Both positive and negative travel-time shifts are seen. The effect is quite small for deeper waves (smaller $\ell$), where the wave approaches the surface almost vertically and hence interacts only weakly with vertical magnetic field, but is greatly enhanced for shallow skippers ($|z_1|\la 5$ Mm).

%%%%%%%%%%%
\begin{table}
 \caption{Spherical harmonic degree $\ell$ for the adopted frequencies and lower turning depths.}
 \label{ell}
 \begin{tabular}{@{}lrrr}
  \hline
   & 3 mHz & 4 mHz & 5 mHz \\
  \hline
  $-3$ Mm & 643 & 936 & 1213 \\
  $-5$ Mm & 521 & 733 & 937 \\
  $-10$ Mm & 341 & 465 & 586 \\
  $-15$ Mm & 274 & 370 & 465 \\
  \hline
 \end{tabular}
\end{table}

\begin{figure*}
\begin{center}
\includegraphics[width=.8\hsize]{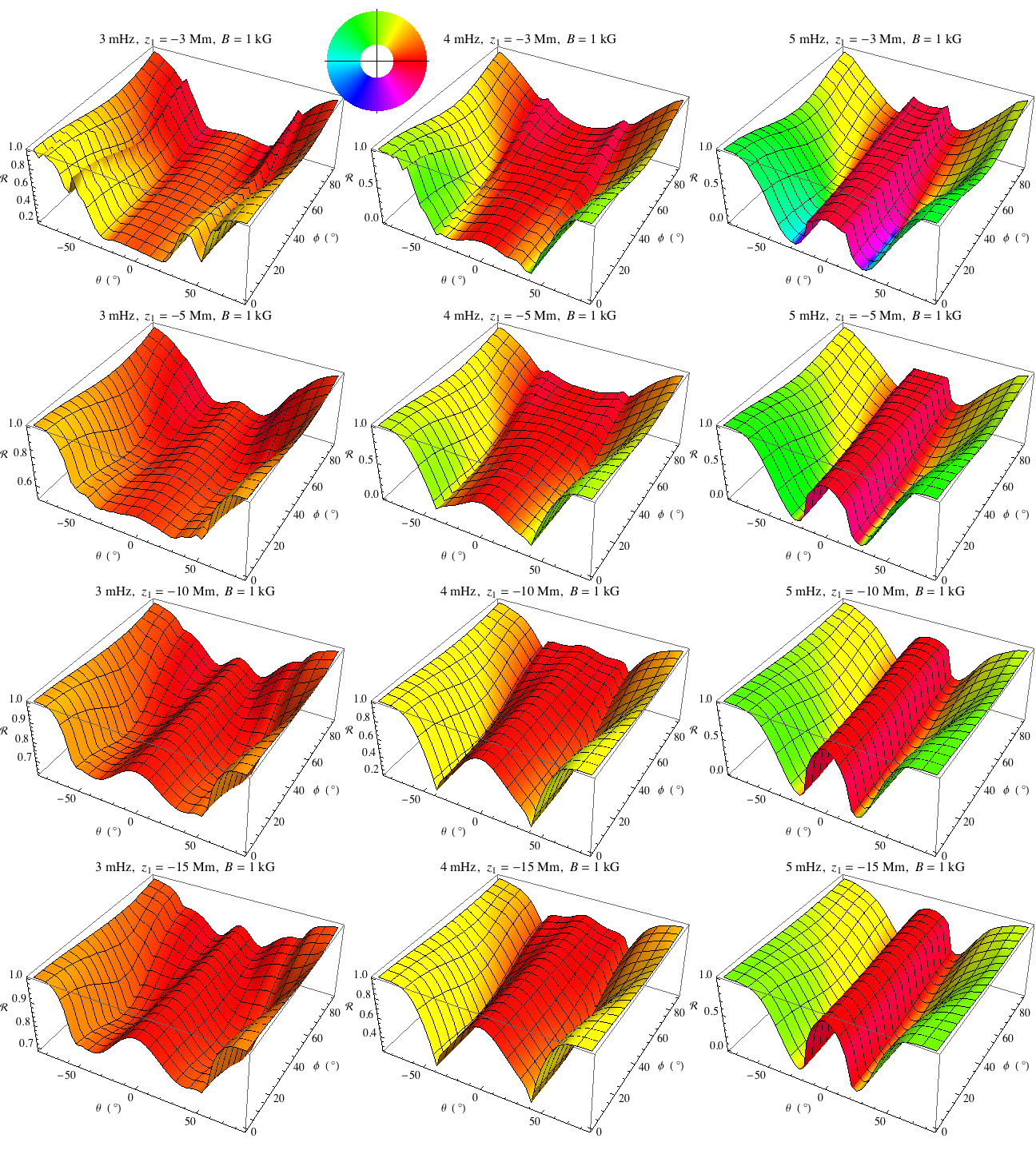}
\caption{Fast wave reflection coefficeint $\mathcal{R}$ (vertical axis) and relative phase jump $\Delta\varphi$ (colour shading) as a function of magnetic field inclination to the vertical $\theta$ and orientation from the $x$-$z$ plane $\phi$. The twelve cases depicted are the combinations of four different turning depths $z_1$ ($-3$ Mm, $-5$ Mm, $-10$ Mm, and $-15$ Mm) and three different frequencies (3 mHz, 4 mHz, and 5 mHz), as labelled. The magnetic field strength is 1 kG throughout. The remainder of the flux is variously lost to the overlying atmosphere as acoustic or magnetic waves, or downward as magnetic waves. The colour shading represents $\Delta\varphi$ on a periodic colour scale (see the legend disk at the top), with red corresponding to $\Delta\varphi=0^\circ=\pm360^\circ$, yellow-green to $90^\circ$, etc. A $4^\circ\times10^\circ$ $\theta$-$\phi$ grid was used throughout.}
\label{fig:B1}
\end{center}
\end{figure*}

\begin{figure*}
\begin{center}
\includegraphics[width=.8\hsize]{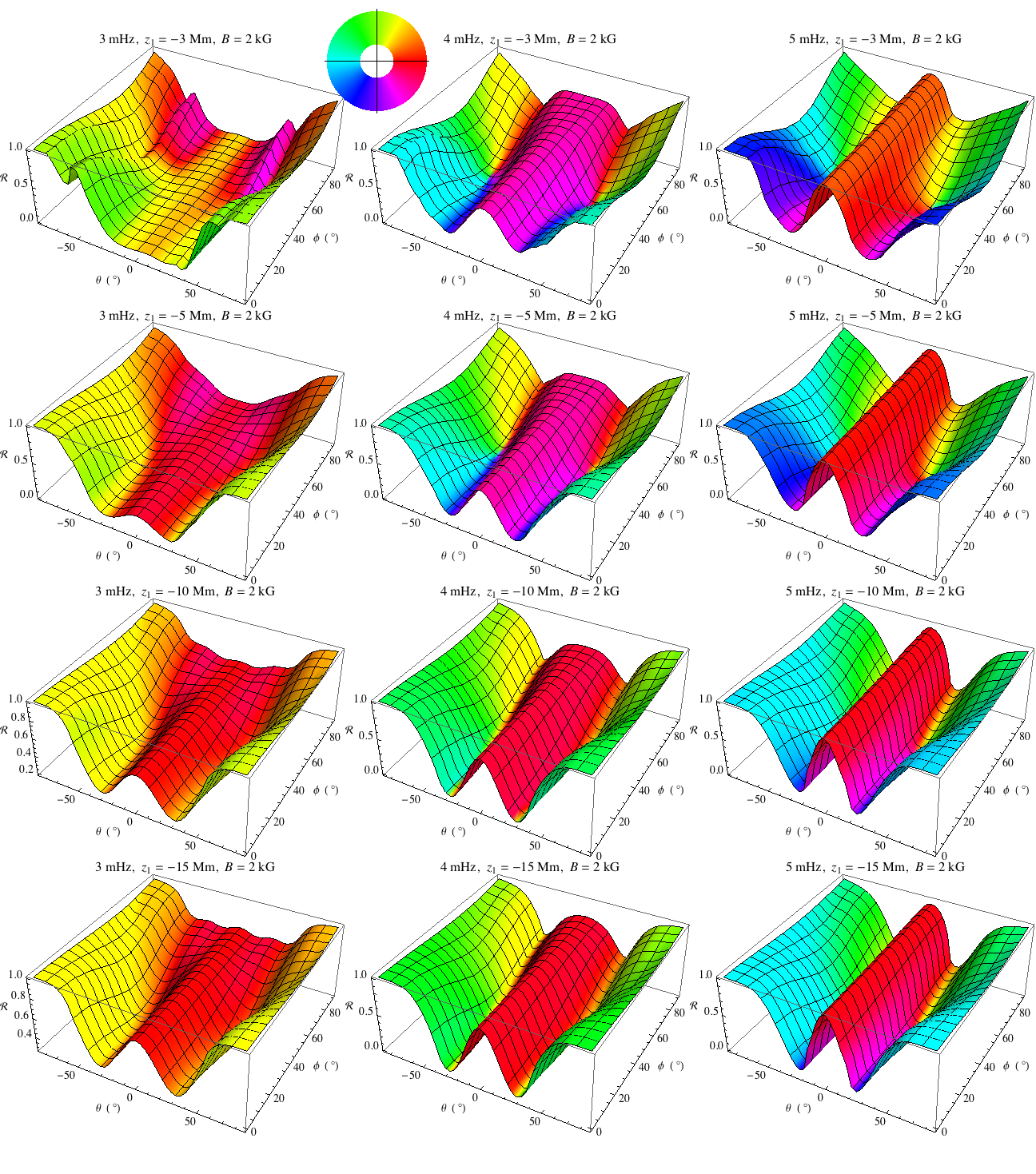}
\caption{As for Figure \ref{fig:B1}, but with 2 kG magnetic field.}
\label{fig:B2}
\end{center}
\end{figure*}

\begin{figure*}
\begin{center}
\includegraphics[width=.8\hsize]{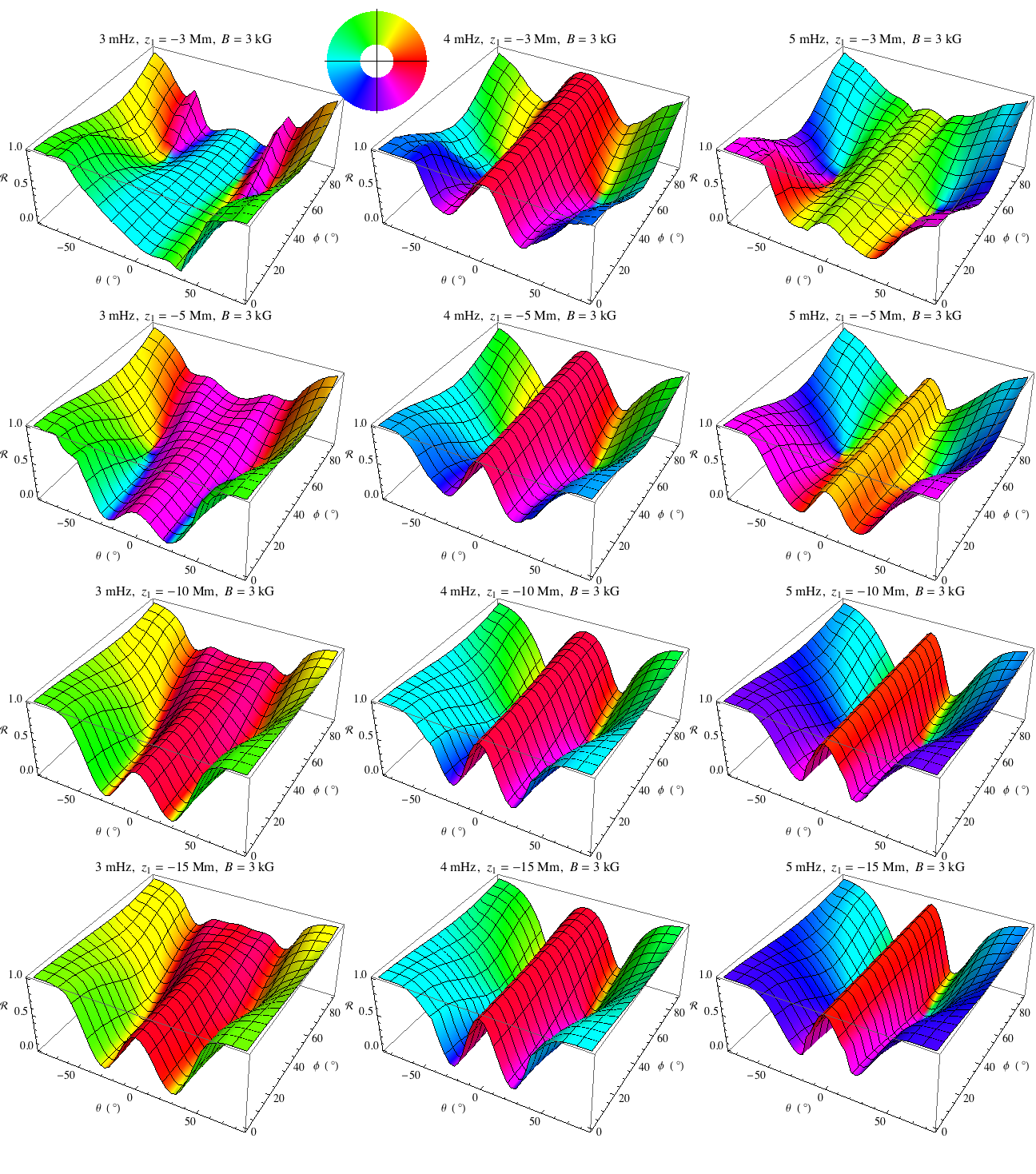}
\caption{As for Figure \ref{fig:B1}, but with 3 kG magnetic field.}
\label{fig:B3}
\end{center}
\end{figure*}

\begin{figure*}
\begin{center}
\includegraphics[width=\hsize]{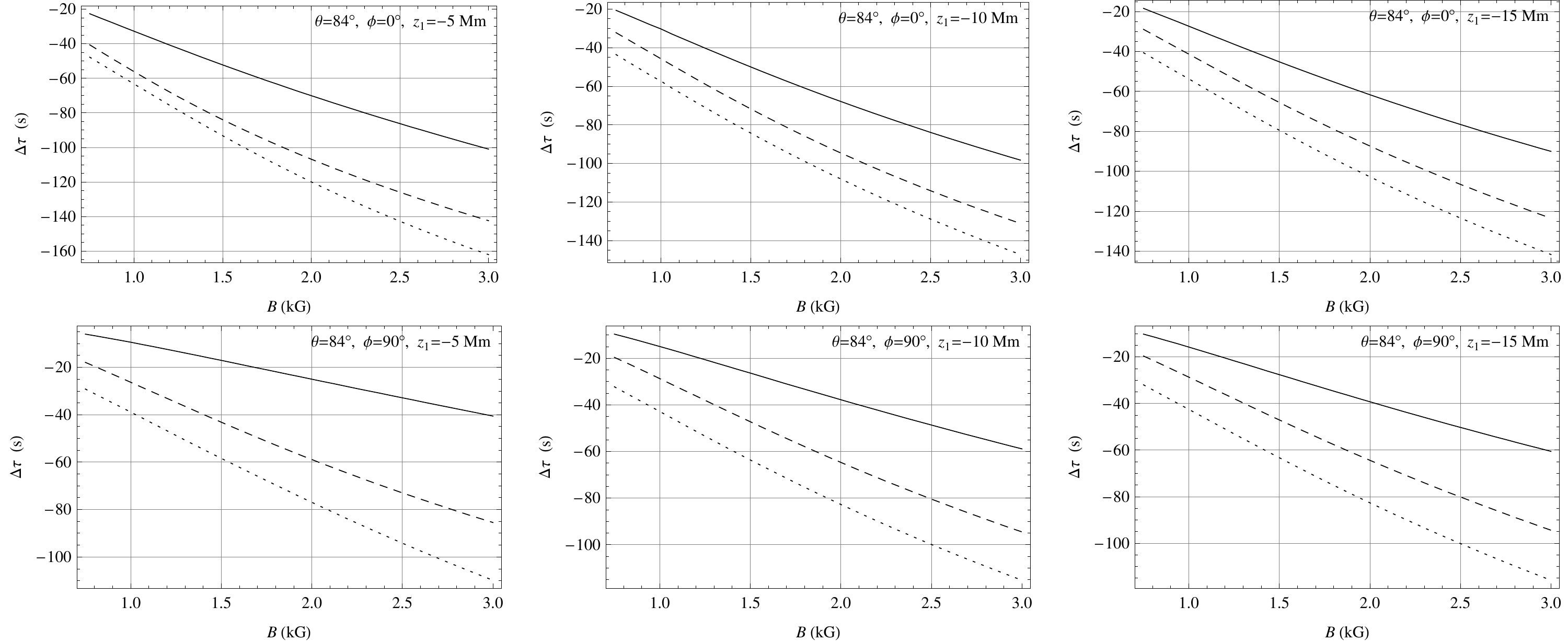}
\caption{Travel-time perturbations $\Delta\tau=-\omega^{-1}\Delta\varphi$ as a function of magnetic field strength (kG) at high field inclination, $\theta=84^\circ$, for turning depths $z_1=-5$ Mm (left column); $z_1=-10$ Mm (centre column); and $z_1=-15$ Mm (right column). The top row corresponds to the 2D case $\phi=0^\circ$, and the bottom row to the perpendicular case $\phi=90^\circ$. The full curve corresponds to a frequency of 3 mHz throughout, dashed to 4 mHz, and dotted to 5 mHz. Highly inclined field in sunspot penumbrae would not normally exceed about 1.5 kG, so travel-time advances of around 1 minute may be expected from these results.}
\label{fig:wide}
\end{center}
\end{figure*}

\subsubsection{3D Inclined Magnetic Field}   \label{sec:3D}
A total of $3\times4\times3=36$ cases are considered, corresponding to three magnetic field strengths $B$, 1 kG, 2 kG, and 3 kG; four lower turning depths $z_1$, $-3$ Mm, $-5$ Mm, $-10$ Mm, and $-15$ Mm; and three wave frequencies, 3 mHz, 4 mHz, and 5 mHz. For $z_1=-3$ Mm we adopt $z_\mathrm{b}=-2$ Mm and $z_\mathrm{s}=-1$ Mm; for $z_1=-5$ Mm we set $z_\mathrm{b}=-4$ Mm and $z_\mathrm{s}=-2$ Mm; and for the other two cases
$z_\mathrm{b}=-8$ Mm and $z_\mathrm{s}=-4$ Mm. The top boundary conditions are applied at $z_\mathrm{t}=2$ Mm throughout. The spherical harmonic degree $\ell$ for each case is listed in Table \ref{ell}. These are typical values in local helioseismology. Note that no attempt is made to place these cases on $p$-mode ridges, in recognition of the ``scattering experiment'' nature of our calculations: a wave is ``fired'' from $z_\mathrm{b}$ and examined when it reflects back to $z_\mathrm{b}$. It is not necessary for it to be a normal mode, as the bottom boundary condition is not homogeneous.\footnote{Strictly speaking, there are no normal modes of the magnetic solar model, as magnetic wave leakage produces pseudo-modes instead, resulting in complex eigenfrequencies or eigen-wavenumbers \citep{cbz,bc97,cc05}. However, the sub-surface acoustic part may be decoupled from the magnetic in a perturbation sense \citep{cally05}, or by using the divergence of the displacement as is done here.}

Figure \ref{fig:B1} shows the fast wave reflection coefficient $\mathcal{R}$ and phase shift $\Delta\varphi$ for $B=1$ kG. The twelve constituent frames correspond to the four turning depths and three frequencies, as labelled, and cover $-84^\circ\le\theta\le84^\circ$, $0\le\phi\le90^\circ$. Of course, results are independent of $\phi$ at $\theta=0^\circ$, and $\Delta\varphi$ is small there in agreement with Figure \ref{fig:B}. However, substantial phase perturbations are found at large inclination $\theta$, especially at 5 mHz. Clearly, fast mode reflection $\mathcal{R}$ is most pronounced at large $\theta$, corresponding to penumbra. This is to be expected, as mode \emph{conversion} is strongest when the wave strikes the magnetic field at $z_\mathrm{eq}$ at large \emph{attack angle} \citep[the angle between the wavevector and the magnetic field lines; see][]{sc06}. At low frequency though, there is still substantial fast wave reflection throughout.

As might be expected, Figures \ref{fig:B2} and \ref{fig:B3} show a generally increasing phase perturbation with increased magnetic field strength. In most cases, maximum phase perturbation is attained at high inclination $\theta$, which is also where the returning fast wave flux is highest. (Of course, not all parts of all frames are relevant to the Sun. Highly inclined field for instance represents penumbra, where field strengths will be somewhat less than 2 kG generally.)

Figure \ref{fig:wide} illustrates that travel-time perturbation $\Delta\tau$ at large field inclination varies essentially linearly with increasing $B$, and also increases in magnitude with increasing frequency. The phase perturbation is somewhat reduced at $\phi=90^\circ$ compared with $\phi=0^\circ$.

It is striking that phase perturbation at high field inclination is relatively insensitive to turning depth $z_1$. It is also notable that the travel-time perturbation is uniformly negative here. Both features are in stark contrast to the vertical field case. The probable cause is that the attack angle is large for all cases in the near-horizontal field, whereas it becomes quite fine for deeper rays in vertical field.

A prominent characteristic of Figures \ref{fig:B1}--\ref{fig:B3}, especially the last two, is the banded structure of the phase perturbation shading, and the associated ``W-shaped'' return flux variation with $\theta$. There is a sharp distinction between $\Delta\varphi$ at small and large inclination, with low fast mode reflectivity $\mathcal{R}$ typically found at the transition region between the two. We focus on the flux first.

\begin{figure}
\begin{center}
\includegraphics[width=0.85\hsize]{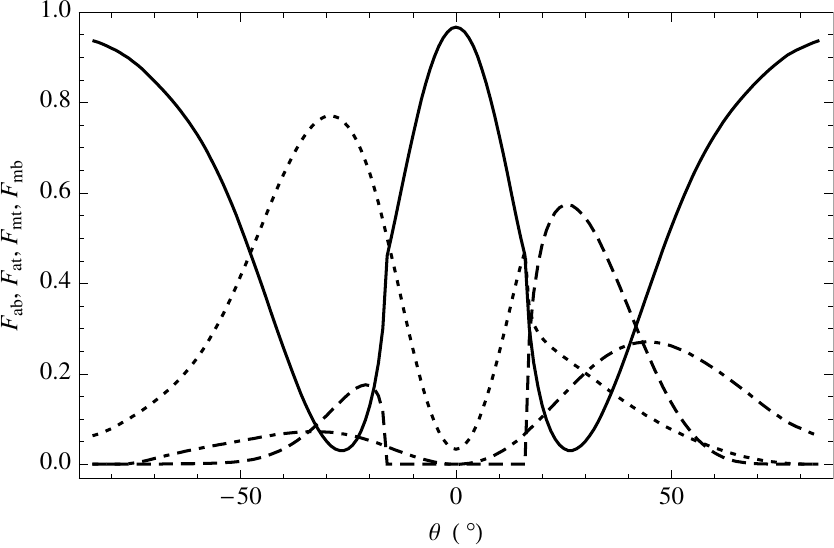}
\caption{Outgoing wave energy fluxes for the 5 mHz case with $z_1=-10$ Mm, $B=2$ kG, $\phi=30^\circ$, corresponding to a slice across the last graph in the third row of Figure \ref{fig:B2} (though with improved $1^\circ$ resolution). Full curve: $F_\mathrm{ab}=\mathcal{R}$ (outward acoustic flux at the bottom -- the ``helioseismic'' field); Dashed curve: $F_\mathrm{at}$ (outward acoustic flux at the top); Dot-dashed curve: $F_\mathrm{mt}$ (Alfv\'enic flux at the top); and Dotted curve: $F_\mathrm{mb}$ (combined slow and Alfv\'en magnetic flux at the bottom). Of course, $F_\mathrm{ab}+F_\mathrm{at}+F_\mathrm{mt}+F_\mathrm{mb}=F_\mathrm{in}=1$, where $F_\mathrm{in}$ is the injected acoustic flux.
}
\label{fig:flux}
\end{center}
\end{figure}

The complex outgoing flux structure with varying field inclination $\theta$ of one of the 5 mHz 3D cases is illustrated in Figure \ref{fig:flux}. The full curve shows the downgoing acoustic flux at $z_\mathrm{b}$, clearly displaying very distinct ``central'' and ``wing'' regions. The reason for this is clear. The upward acoustic flux (dashed curve) turns on sharply once $|\theta| > \arccos(\omega/\omega_c)=16.2^\circ$ in this case. The upward acoustic losses are maximal where the attack angle is smallest, around $30^\circ$ here, in agreement with the complex eigenmode calculations of \cite{cc05}. They are small for negative $\theta$ because the attack angle is large there. On the other hand, the downward magnetic losses (dotted curve) display the opposite behaviour. This is because they result from mode transmission on the downward leg of the fast ray, where now attack angle is smallest around $-30^\circ$ and large at positive $\theta$. The upgoing Alfv\'en flux (dot-dashed curve) also displays an inclination asymmetry. So, in this and many other cases, the ``centre/wing'' dichotomy in $\mathcal{R}$ is easily understood from flux graphs, and is in complete accord with the generalized ray-theoretic insights of \cite{sc06}. This angle-dependent ``absorption'' of helioseismic waves by sunspots also explains the Hankel analysis results of \cite{braun95} \citep[see also][where calculated complex eigenvalues are used to reproduce the Hankel absorption and phase shift data quite accurately]{ccb03,cccbd05}. The weak dependence of ``absorption coefficient'' $\alpha=1-\mathcal{R}$ on $\phi$ was previously noted by \cite{cc05}.

The reason for the banded phase structure is less obvious, in the current absence of a simple generalized ray-theoretic description of phase jumps at mode conversion points. However, the behaviour is very reminiscent of the ``central dips'' seen in phase jumps across mode conversion levels in Figure 8 of \cite{cally08}. We therefore tentatively attribute the effect to mode-conversion rather than turning point phase jumps, which the isothermal model suggests should be relatively insensitive to field inclination.

%%%%%%%%%%%%%%%%%%%%%%%%%%%%%%%%%%%%%%

\begin{figure*}
\begin{center}
\includegraphics[width=\hsize]{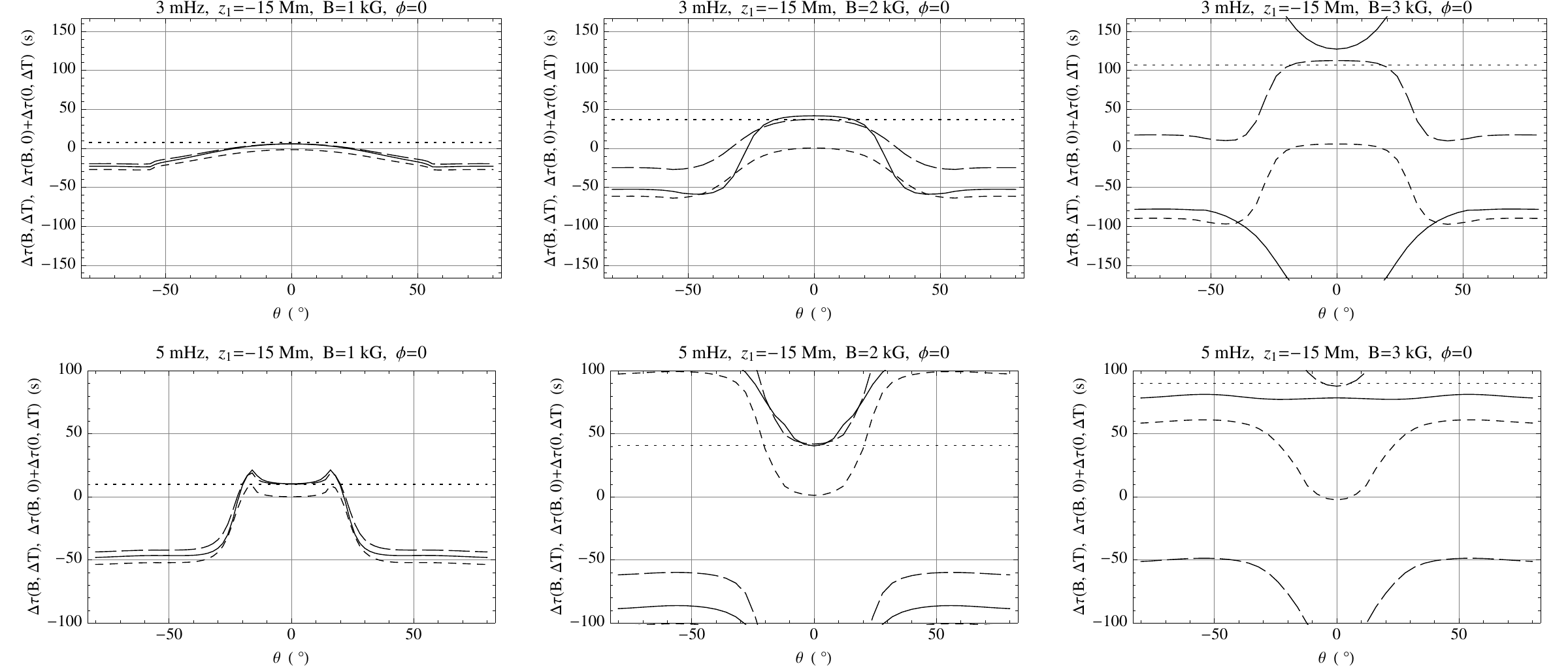}
\caption{Travel-time perturbations as a function of magnetic field angle $\theta$ with $\phi=0^\circ$ for frequencies of 3 mHz (top row) and 5 mHz (bottom row) and 1 kG magnetic field (left column), 2 kG (centre column) and 3 kG (right column). The full curve represents $\Delta\tau(B,\Delta T)$, the long-dashed curve $\Delta\tau(B,0)+\Delta\tau(0,\Delta T)$, short-dashed $\Delta\tau(B,0)$, and dotted $\Delta\tau(0,\Delta T)$.  The $\Delta\tau$ range plotted corresponds to $-180^\circ\le\Delta\varphi\le180^\circ$, and in several cases displays a $360^\circ$ ``wrap-around''.}
\label{fig:both}
\end{center}
\end{figure*}

\subsection{Combined Magnetic and Thermal Anomalies}   \label{sec:comb}
Let us now explore the combined effects of the magnetic and thermal anomalies. The travel-time perturbation due to both magnetic and thermal anomalies may be represented by $\Delta\tau(B,\Delta T)$. Is this well approximated by the sum of the individual effects $\Delta\tau(B,0)+\Delta\tau(0,\Delta T)$? Figure \ref{fig:both} makes it clear that the answer is ``yes'' for weak magnetic field (1 kG, left column), where the full and long-dashed curves are very close, but a resounding ``no'' at 3 kG, unless the field is nearly vertical. At 2 kG, we have a qualified ``yes'' for $|\theta|\la 25^\circ$.

\begin{figure*}
\begin{center}
\includegraphics[width=\hsize]{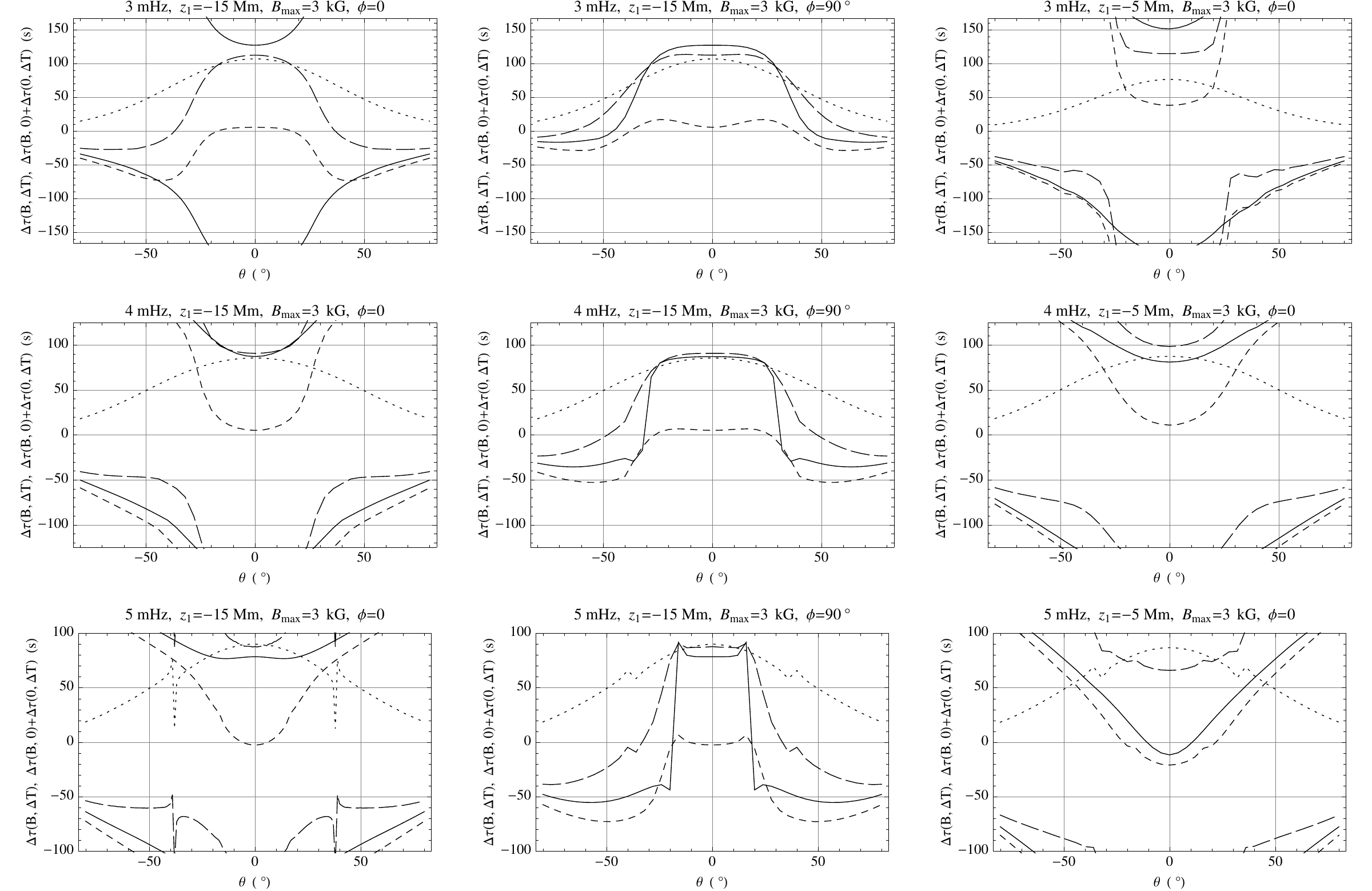}
\caption{Travel-time perturbations in the MMMS model as a function of magnetic field angle $\theta$ with $z_1=-15$ Mm and $\phi=0^\circ$ (left column), $z_1=-15$ Mm and $\phi=90^\circ$ (centre column), and $z_1=-5$ Mm and $\phi=0^\circ$ (right column), for $B=3\,\exp[-(\theta/90^\circ)^2]$ kG and three frequencies: 3 mHz (top), 4 mHz (centre), and 5 mHz (bottom). The full curve represents $\Delta\tau(B,\Delta T)$, the long-dashed curve $\Delta\tau(B,0)+\Delta\tau(0,\Delta T)$, short-dashed $\Delta\tau(B,0)$, and dotted $\Delta\tau(0,\Delta T)$. The $\Delta\tau$ range plotted corresponds to $-180^\circ\le\Delta\varphi\le180^\circ$, and in several cases exhibits a $360^\circ$ ``wrap-around''.}
\label{fig:bothG}
\end{center}
\end{figure*}

Of course though, highly inclined field (penumbra) typically does not exceed 2 kG, so the large $\theta$ curves at $B=3$ kG are moot. We incorporate the tendency for field strength to decrease with field inclination by setting $B=B_\mathrm{max}\,\exp[-(\theta/90^\circ)^2]$, with $B_\mathrm{max}=3$ kG, which yields 1.1 kG at $\theta=90^\circ$. Figure \ref{fig:bothG} then displays rather complex behaviour. Negative travel-time shifts appear characteristic of the penumbra, though positive shifts are typically seen in the umbra (small $|\theta|$), in accord with the observations of \citet{cr07}. In the $\phi=0^\circ$ cases (2D), this is due to the $360^\circ$ ambiguity in $\Delta\varphi$, both for deep (left column) and shallow (right colum) skipping waves. For the transverse 3D case ($\phi=90^\circ$, centre column), all solutions displayed stay within the principal window. For the most part, the assumption of additive thermal and magnetic effects is rather poor in the penumbra, and only sometimes good in the umbra. In fact, it appears that $\Delta\tau(B,\Delta T)\approx\Delta\tau(B,0)$ is a better approximation than $\Delta\tau(B,\Delta T)\approx\Delta\tau(B,0)+\Delta\tau(0,\Delta T)$ in penumbrae.

\begin{figure*}
\begin{center}
\includegraphics[width=\hsize]{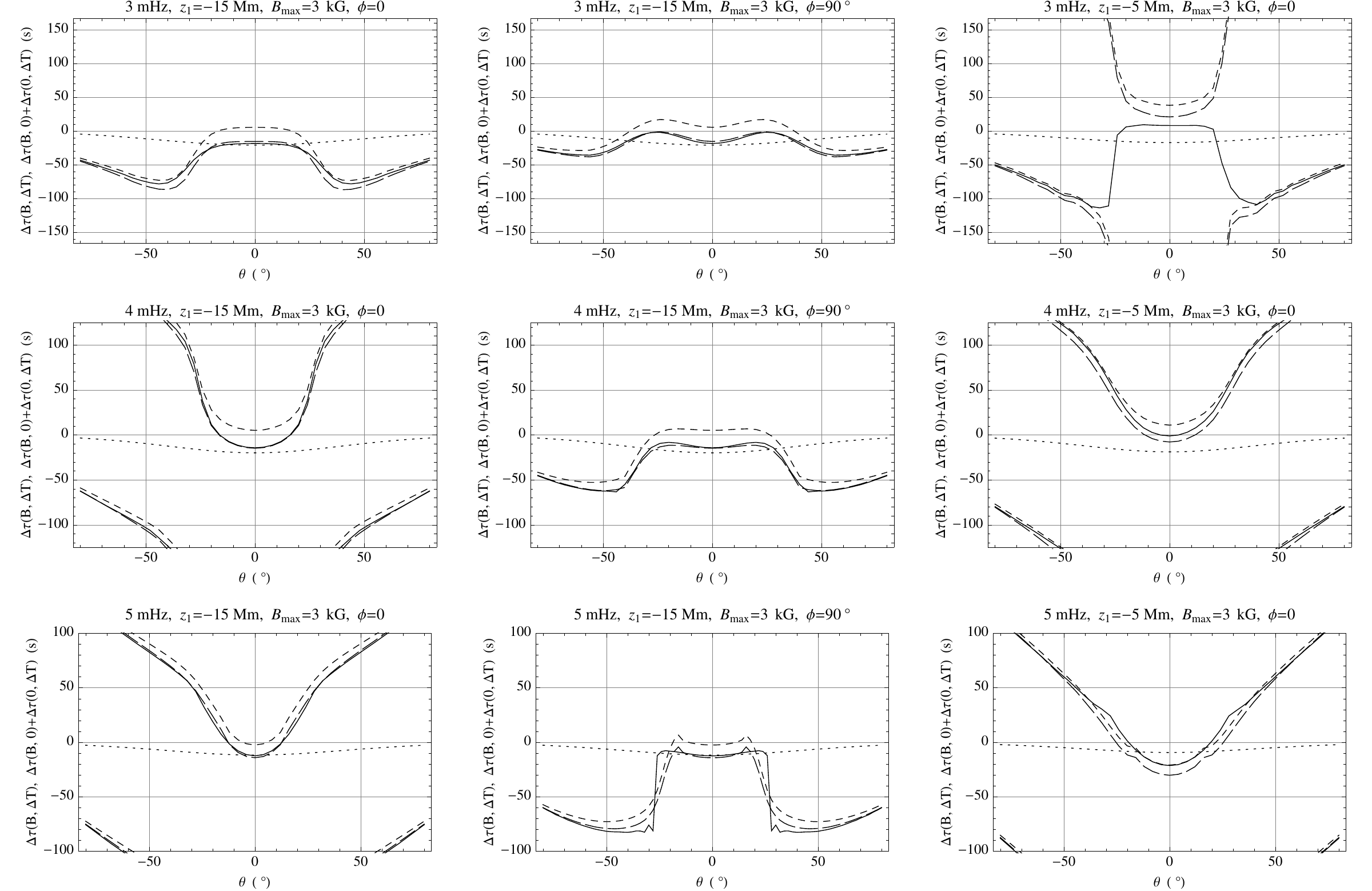}
\caption{Same as Figure \ref{fig:bothG}, but for a thermal model with reduced density as well as temperature, thereby producing negative non-magnetic travel-time perturbations (dotted curves).}
\label{fig:AS}
\end{center}
\end{figure*}

Finally, the calculations of Figure \ref{fig:bothG} are repeated with an alternate heuristic thermal model in which the density is not held fixed. In fact, the scale height $H$ \emph{decreases} in the near-surface layers as they are cooled, sufficiently fast that $\omega_c=(c/2H)\sqrt{1-2H'}$ actually increases. This results in shorter travel paths and \emph{negative} travel time shifts in the non-magnetic case, despite the reduced sound speed. Figure \ref{fig:AS} illustrates the results. Again, it is clear that direct magnetic effects dominate $\Delta\tau$ in penumbrae, but the umbral shifts are predominantly thermal. However, we might also conclude that $\Delta\tau(B,\Delta T)\approx\Delta\tau(B,0)+\Delta\tau(0,\Delta T)$ does rather better in this model, probably because the magnitude of $\Delta\tau(0,\Delta T)$ is typically much smaller.

%%%%%%%%%%%%%%%%%%%%%%%%%%%%%%%%%%%%%%%%%%%%%%%%%%%%%
\section{DISCUSSION}

The analysis in \cite{cally08} largely proceeds by following a fast mode ray through the surface layers as it first propagates upward, refracts from the Alfv\'en speed gradient, and finally propagates downward. The ray is used as a ``sampler'' of the known exact wave field in the 2D isothermal case: \emph{i.e.}, determining the \emph{exact} phase at the \emph{ray-theoretic} position $\mathbf{x}(t)$ at the time $t=T$ when the ray returns to $z=z_\mathrm{b}$. One may also compare this to the ray-theoretic phase $S(T)$ calculated along with the ray path. We have chosen not to apply that procedure here, largely because the ray integration is unreliable in realistic solar models near the surface, for reasons associated with the small vertical length scales, and the definition of the acoustic cutoff frequency \citep[see][]{sc06}. On the other hand, the definition of $\delta\varphi$ in Equation (\ref{R-delta}) samples phase ``in place'', \emph{i.e.}, $\delta\varphi=-\arg(C_-/C_+)$ rather than $-\arg(C_-/C_+)+ k\,x(T)-\omega\,T$. It pays no heed to the differing ray paths and travel times. Nevertheless, the ``in place'' $\delta\varphi$ is made useful by comparison to the nonmagnetic phase difference between downgoing and upgoing acoustic waves at $z_\mathrm{b}$, \emph{i.e.}, $\Delta\varphi=\delta\varphi-\delta\varphi_0$. This is consistent with the standard local helioseismic perspective, where comparison is always made to quiet Sun values. 

A clear lesson to be drawn from all the magnetic-only cases presented in Section \ref{sec:magonly} is that phase perturbation $\Delta\varphi$ is most pronounced for highly inclined magnetic field, characteristic of sunspot penumbrae. The significance of this correspondence is enhanced by the maximal fast mode reflection coefficient $\mathcal{R}$ at large inclination. Therefore, penumbra (and perhaps plage with its predominantly horizontal overlying canopy field) may be expected to be most reflective and also most productive of phase astigmatism. This appears to be consistent with observational evidence from TD and HH \citep{cr07,lb05,sbcl05}.

The travel-time perturbations presented here, in Figure \ref{fig:B} (umbra) and Figure \ref{fig:wide} (penumbra), and indirectly Figures \ref{fig:B1}--\ref{fig:B3}, should be understood as due to direct magnetic effects on the waves, through mode conversion, transmission, and fast wave reflection.

In addition to these effects though, the magnetic field of sunspots significantly alters the near-surface thermal structure, and therefore the sound speed. This is addressed in Section \ref{sec:comb} with an ad hoc thermal model (MMMS). We find a definite tendency for negative travel-time shifts of around a minute in highly inclined (and comparatively weak) penumbral field. The behaviour in more vertical (and stronger) umbral field is rather complex, often associated with the $360^\circ$ phase ambiguity, but is typically positive if the window $-180^\circ\le\Delta\varphi\le180^\circ$ is assumed. This raises a further doubt about the wisdom of interpreting $\Delta\tau=-\omega^{-1}\Delta\varphi$ as a true ``travel-time'' perturbation: the $360^\circ$ wrapping can turn an actual large negative travel-time shift into an apparent positive perturbation quite easily. It is tempting to hypothesize that this may be responsible for the positive umbral shifts reported by \citet{cr07}. Alternate plausible thermal models may produce quite different thermal shifts (figure \ref{fig:AS}), but the general conclusions are unchanged.

Based on the results illustrated in Figures \ref{fig:bothG} and \ref{fig:AS}, a crude rule-of-thumb appears to be that travel-time perturbations in umbrae are predominantly thermal, whereas in penumbrae they are mostly magnetic. This is a neat explanation for the umbra/penumbra dichotomy identified by \citet{cr07}. The explanation for this behaviour is clear: for heliosesimic rays propagating steeply to the surface, the direct interaction with the near-vertical magnetic field is small, resulting in substantial mode \emph{transmission} (acoustic-to-acoustic) across the equipartition layer, almost as if the field were not there. On the other hand, mode \emph{conversion} (acoustic-to-magnetic) dominates for highly inclined penumbral field. The resulting atmospheric fast wave quickly reflects back downward to rejoin the helioseismic wave field, but with a very different phase (and timing) to the non-magnetic case.

Current local helioseismic techniques are particularly adept at measuring temperature, as well as flow speeds. However, they are easily confused by the intrinsic magnetic perturbations. It is hoped that the results presented here will point the way to interpreting observed local-helioseismic data in sunspots. Further forward modelling based on more or less realistic magnetic and thermal sunspot models \citep{cgd08,mc08,mhc08} is also warranted.

%%%%%%%%%%%%%%
\section*{Acknowledgement}
Thanks to Hamed Moradi and Charlie Lindsey for asking the right questions.

%%%%%%%%%%%%%%%%%%%%%%%%%%%%%%%%%%%%%%%%%%%%%%%%%%%%%
%%%%%%%%%%%%%%%%%%%%%%%%%%%%%%%%%%%%%%%%%%%%%%%%%%%%%
%%%%%%%%%%%%%%%%%%%%%%%%%%%%%%%%%%%%%%%%%%%%%%%%%%%%%%%%%%%%%%%%%%%%%%%%%%%%%
%%%%%%%%%%%%%%%%%%%%%%%%%%%%%%%%%%%%%%%%%%%%%%%%%%%%%%%%%%%%%%%%%%%%%%%%%%%%%%%%
%%%%%%%%%%%%%%%%%%%%%%%%%%%%%%%%%%%%%%%%%%%%%%%%%%%%%%%%%%%%%%%%%%%%%%%%%%%%%
%%%%%%%%%%%%%%%%%%%%%%%%%%%%%%%%%%%%%%%%%%%%%%%%%%%%%
%%%%%%%%%%%%%%%%%%%%%%%%%%%%%%%%%%%%%%%%%%%%%%%%%%%%%
%%%%%%%%%%%%%%%%%%%%%%%%%%%%%%%%%%%%%%%%%%%%%%%%%%%%%

%%%%%%%%%%%%%%%%%%%%%%%%%%%%%%%%%%%%%%%%%%%%%%%%%%%%%%%%%%%%%%%%%%%%%%
\appendix
\section{Nonmagnetic and WKB Solution} \label{app:A}

In the nonmagnetic case, the wave equation may be written in the form 
\begin{equation}
\Psi''=\omega^2\,Q(z)\,\Psi\, , \label{WKBde}
\end{equation}
where 
\begin{equation}
\omega^2\,Q(z) = -\left[\frac{\omega^2-\omega_c^2}{c^2}+
\left(\frac{N^2}{\omega^2}-1\right)k_x^2  \right]\, ,
\end{equation}
and $N$ is the {\bv} frequency. The acoustic cutoff frequency is specified by $\omega_c^2=(c^2/4H^2)(1-2H')$, where $H$ is the density scale height. (For numerical reasons associated with difficulties in calculating $H'$ smoothly and reliably in tabulated solar models, this is not the formulation we use for numerical solution in the nonmagnetic case, though mathematically it is equivalent to it.)\phantom{.} Equation (\ref{WKBde}) is in convenient form for WKB solution, with $\omega^{-1}$ taking the role of the perturbation parameter. The WKB solution in the acoustic cavity, where $Q$ is negative, is
\begin{equation}
\Psi \sim f_\pm = (-Q(z))^{-1/4}
\exp\left[\pm \i\,\omega\int^z \sqrt{-Q(z')}\,\rd z' \right]
\end{equation}
as $\omega\to\infty$. See \cite{cally08} for further details. A linear combination of $f_+$ and $f_-$ may be used to fit the numerical magnetic solution $\Psi$ at sufficient depth that the magnetic and acoustic waves have decoupled. This identifies the upgoing and downgoing components.

\end{document}